\documentclass[useAMS,usenatbib]{mn2e}

\usepackage{natbib}
\usepackage{graphicx}
\usepackage{times}
\title[Multi-wavelength Diameters of Miras and Semiregulars]
{Multi-wavelength Diameters of Nearby Miras and Semiregulars}
\author[M.J. Ireland et al.]{M.J. ~Ireland,\thanks{E-mail: mireland@physics.usyd.edu.au} P.G. ~Tuthill,
  T.R. ~Bedding, J.G. ~Robertson and A.P. ~Jacob \\
School of Physics, University of Sydney, NSW 2006, Australia}

\begin{document}


\pagerange{\pageref{firstpage}--\pageref{lastpage}} \pubyear{2004}

\maketitle

\label{firstpage}

\begin{abstract}
We have used optical interferometry to obtain multi-wavelength visibility
curves for eight red giants over the wavelength range 650--1000 nm. The
observations consist of wavelength-dispersed fringes recorded with MAPPIT
(Masked APerture-Plane Interference Telescope) at the 3.9-m
Anglo-Australian Telescope. We present results for four Miras (R~Car,
$o$~Cet, R~Hya, R~Leo) and four semi-regular variables (R~Dor, W~Hya,
L$_2$~Pup, $\gamma$~Cru).  All stars except $\gamma$~Cru show strong
variations of angular size with wavelength.  A uniform-disk model was found
to be a poor fit in most cases, with Gaussian (or other more tapered
profiles) preferred.  This, together with the fact that most stars showed a
systematic increase in apparent size toward the blue and a
larger-than-expected linear size, even in the red, all point toward
significant scattering by dust in the inner circumstellar environment.
Some stars showed evidence for asymmetric brightness profiles, while
L$_2$~Pup required a two-component model, indicating an asymmetrical
circumstellar dust shell.
\end{abstract}

\begin{keywords}
techniques: interferometry -- stars: AGB and post-AGB
\end{keywords}

\section{Introduction}

Late-type giants have particularly extended atmospheres, which makes
it difficult to define a particular value for the stellar
diameter \citep{Baschek91,Scholz01}.  
Even within the narrower context of observed intensity diameters, 
their complicated center-to-limb brightness profiles and strong dependence on  
bandpass from absorption in atmospheric layers leaves no simple 
``continuum diameter'' to be measured.

To truly characterise the brightness distribution of a single
late-type giant, one would thus like the complete stellar intensity profile
at all wavelengths. 
Since all of these stars pulsate and some show long-term cycle-to-cycle 
variations in diameter 
(eg \citealt{Tuthill95}), as expected from models \citep{Hofmann98}, this information 
must be obtained simultaneously 
and the observations then repeated at many different pulsation phases.

There is a significant history of multi-wavelength angular diameter
measurements of red giants, beginning with \citet{Bonneau73}. 
Unfortunately, difficulties in calibration and the availability of 
only a few selected bandpasses has meant that spectral coverage of 
apparent diameter measurements has been limited. 
In this paper, we present simultaneous multi-wavelength measurements of 
the apparent sizes of 8 giants using aperture masking interferometry.
With a maximum baseline of 3.89\,m, our visibility data (related to the
object intensity profile by the Fourier transform) were able to 
resolve all the stars in this sample.

\section{Observations and Data Reduction}

MAPPIT (Masked APerture-Plane Interference Telescope) was an
aperture masking system set up at the coud\'{e} focus of the
Anglo-Australian Telescope (AAT). 
The pupil of the AAT was re-imaged onto a mask, with starlight
subsequently passing through a prism and cylindrical lens.
The optical setup resulted in the image formed at the detector 
having high-resolution spatial information (fringes) in one 
direction, and a spectrum of the source in the other.
Details of the experiment can be found in \citet{Bedding94}, with 
the only significant change being the replacement of the IPCS
detector with a CCD, accompanied by a change in wavelength
coverage to the range 650\,nm -- 1\,$\mu$m. Previous work with this
experimental setup can be found in \citet{Jacob03}.

We used MAPPIT to record wavelength-dispersed fringes from each of three
masks: two five-hole linear masks with maximum baselines of 3.2 and 
3.8 m, and one slit mask that utilised the full 3.89 m aperture of the
AAT\@. 
Hole/slit widths of 9, 15 and 18\,cm,  as projected onto the primary 
mirror, were available. 
Data from the wider holes/slits were more influenced by atmospheric noise,
while the narrower holes/slits had a narrower effective bandpass,
greater resistance against atmospheric noise but inevitably lower transmission.
The linear nature of the masks results in only a one-dimensional 
brightness profile being recorded, however an image rotator enabled
different orientations on the sky to be sampled. 

The data analysed here were collected on the nights of February 8/9, 2001, with
selected known parameters for the target
stars given in Table~\ref{tblStarParams} and observational parameters presented in
Table~\ref{tblObservations}. Seeing on these nights was above average
for the AAT, ranging from 0.8 to 1.3 arcsec.
A data set for one target star at one position angle consisted of 2--4
sets of 100--200 frames, with identical 
observations of an unresolved or barely-resolved calibrator
star. Exposure times ranged from 10 to 40\,ms. Calibration of visibility
amplitudes was done by dividing the observed squared-visibilities of the target
star by those of the calibrator. The effects of seeing were not
identical between source and calibrator star, due to differences of
up to 10 degrees in position on the sky, as well as temporal changes
in seeing over the approximately 10 minutes required to interleave
observations. To account for this in model fitting, a self-consistent 
scaling factor (generally between 0.8 and 1.2 in $V^2$) was applied, 
and the error bars compensated for miscalibration by comparing
consecutive observations of different calibrator stars. Calibration of
the differential phase of the fringes of target stars across the system
bandwidth was not possible, due to
uncertainties in system dispersion caused primarily by misalignment
errors and differing paths of calibrator and target stars through
dispersive optics. Thus, the only phase information used was that
contained in the closure phases. Errors not included in the following
section are a 1\% uncertainty in both the wavelength scale and the
detector pixel scale.                                  

\begin{table}
\caption{Details of the stars presented in this paper. Phases at the time of 
the MAPPIT observations were estimated from visual observations.} 
\begin{tabular}{@{}lrrrr@{}}
  \hline
   Name     & Pulsation & Period & {\em Hipparcos} \\    
            & phase     & (d)    & Parallax (mas) \\     
 \hline
R~Car      & 0.35            & 309 & $ 7.4 \pm 0.8$\rlap{$^b$} \\   
$o$~Cet    & 0.34            & 332 & $ 7.8 \pm 1.1$\rlap{$^a$} \\    
R~Hya      & 0.62            & 389 & $ 8.4 \pm 1.0$\rlap{$^b$} \\   
R~Leo      & 0.91            & 310 & $12.2 \pm 1.4$\rlap{$^b$} \\  
R~Dor      & 0.83\rlap{$^c$} & 332 & $17.0 \pm 0.6$\rlap{$^b$} \\  
W~Hya      & 0.44            & 361 & $12.9 \pm 1.0$\rlap{$^b$} \\  
L$_2$~Pup  & 0.58            & 140 & $16.5 \pm 1.3$\rlap{$^a$} \\ 
$\gamma$~Cru & --            & 13--16  & $37.1 \pm 0.7$\rlap{$^a$} \\      

\hline
\end{tabular}
\newline $^a$ \citet{Perryman97}
\newline $^b$ {\em Hipparcos} data reprocessed by \citet{Pourbaix03} with
a revised chromaticity correction and a more stringent criterion for
accepting a variability-induced-mover solution.
\newline $^c$  R~Dor has two pulsation periods: see text.
\label{tblStarParams}
\end{table}

\begin{table}
\caption{Details of the observations presented in this paper. All
  observations
  occurred within twenty-four hours of the Julian Date 2451949.5
  (February 8/9, 2001). The (L) and (S)
  designations represent the long (3.8\,m) and short (3.2\,m) 5-hole masks
  respectively. Dimensions are as projected onto the primary mirror, 
  and position angles are given as angles North through 
  East. } 
\begin{tabular}{@{}lllrr@{}}
  \hline
   Name     & Calibrators & Mask & Hole/slit & Position \\
            &             & type & size (cm) & angle ($^\circ$)  \\
 \hline
   R~Car    & HR~4050      & Slit & 18 & 219 \\
   $o$~Cet  & $\alpha$~Cet & Slit & 15 & 225 \\
   $o$~Cet  & $\alpha$~Cet & Hole(S) & 18 & 274 \\
   R~Hya    & 2~Cen, $\pi$~Hya        & Slit & 18 & 72 \\
   R~Hya    & $\gamma$~Hya, $\pi$~Hya & Slit & 18 & 123 \\   
   R~Leo    & $\alpha$~Leo & Slit & 9 & 209 \\
   R~Leo    & $\alpha$~Leo & Slit & 9 & 305 \\
   R~Leo    & $\alpha$~Leo & Hole(S) & 15 & 280 \\
   R~Leo    & $\alpha$~Leo & Hole(S) & 15 & 276 \\
   R~Dor    & $\gamma$~Ret & Hole(S) & 18 & 175 \\
   R~Dor    & $\gamma$~Ret & Hole(S) & 18 & 260 \\
   W~Hya    & $\theta$~Cen & Slit & 18 & 252 \\
   W~Hya    & 2~Cen, $\gamma$~Hya, & Slit & 18 & 120 \\
            & $\pi$~Hya & & & \\
   L$_2$~Pup    &  $\pi$~Pup& Hole(S) & 18 & 178 \\
   L$_2$~Pup    &  $\pi$~Pup& Slit    & 9 & 209 \\
   $\gamma$~Cru & $\alpha^1$~Cru & Hole(L) & 9 & 87 \\

\hline
\end{tabular}
\label{tblObservations}
\end{table}

The calibrators for each target star are given in Table~\ref{tblObservations}. 
Corrections were made for partially resolved calibrators,
where their angular diameters were estimated from the linear diameters given
in \citet{Dumm98} and their
{\em Hipparcos} parallaxes. This correction was largest for the bright
calibrator 2~Cen, with expected angular size 14.7 mas, which was always
used in conjunction with other calibrators to minimise errors.

\subsection{Coherent Averaging}

It is well known in long-baseline optical interferometry that in the
photon-starved regime, the signal-to-noise ratio of interferometric
data depends primarily on the coherent integration time
\citep{Quirrenbach94}. Although aperture masking does not permit fringe
tracking, in post-processing one can use estimates of the atmospheric
phase offsets between holes to find a phasor multiplier for each complex
visibility. This removes the atmospheric phase noise in the fringes, allowing
 coherent averaging of visibilities with an
associated increase in signal-to-noise \citep{Marson97}. 

To illustrate the benefits of coherent averaging, consider the simple
noise model of a single measurement of complex visibility:
\begin{equation}
 V_m  = V + \sigma(Z_r + iZ_i).
\label{eqn:NoiseDefn}
\end{equation}
Here, $V$ is the true visibility, $Z_r$ and $Z_i$ are independent standard
normal distributions and $\sigma$ represents the noise in a single
measurement.  For MAPPIT data, $\sigma$ was often dominated by CCD readout
noise.  When fringe power, or square visibility modulus is averaged, the
resultant incoherently averaged signal-to-noise in $|V|^2$ is:
\begin{equation}
  SN_I = \frac{N_f|V|^2}{2\sigma\sqrt{N_f\sigma^2 + N_f|V|^2}},
\label{eqn:SNIncoherent}
\end{equation}
where $N_f$ is the number frames of data with other symbols as before.  If
coherent averaging is used, the complex visibilities are averaged and the
resultant coherently averaged signal-to-noise is:
\begin{equation}
  SN_C = \frac{N_f^2|V|^2}{2N_f\sigma\sqrt{\sigma^2 + N_f|V|^2}}.
\label{eqn:SNCoherent}
\end{equation}
In the low signal-to-noise regime, coherent averaging gives a maximum
increase in signal-to-noise of $\sqrt{N_f}$ for $|V|/\sigma \ll 1$. For
example, this increase in signal-to-noise is as large as 20 for 400 frames
of data.

It was possible to estimate the phase offsets between holes for MAPPIT data
taken with a hole mask by using a combination of baseline (or phase)
bootstrapping and wavelength bootstrapping (see, for example,
\citealt{Armstrong98}). For coherent averaging to work without the
possibility of miscalibration errors, these phase offsets needed to be
estimated to an accuracy of much better than 1 radian. This in turn meant
that the single-frame signal-to-noise had to be $\gg 1$ for more
visibilities than the number of atmospheric degrees of freedom. This
condition was certainly met in the case of the stars considered here, which
had high signal-to-noise for short baselines in pseudo-continuum
bands. Estimating the phase offsets in a holistic way (rather than simply
choosing a few short baselines and wavelengths from which to bootstrap)
required a knowledge of the closure phases at all wavelengths, as well as
differential phase offsets between wavelength channels. This knowledge was
developed iteratively by starting with a symmetrical model, solving for the
the phase offsets for each frame of data, coherently averaging the complex
visibilities, and then producing a new model.

This process is similar to spectral line self-calibration as used in radio
astronomy \citep{Taylor99}, but with a few key differences. The complex
telescope gains as used in radio astronomy here become phasors of the form
$e^{2\pi il_{j}/\lambda_{k}}$, where $\lambda_{k}$ is the wavelength of
spectral channel $k$, and $l_{j}$ is an optical path offset for the $j$th
hole. This replacement of phases $\phi_{j}$ with delays $l_{j}$ was
required because of the large total bandwidth of MAPPIT in combination with
atmospherically induced delay errors of up to several microns. Also, no
information could be gained from the time dependence of atmospheric delays
because there was no correlation between delays in successive frames of
data. This is easily explained by the large CCD readout times for MAPPIT
(10--40 ms exposures and 650 ms readout time). Lastly, the self-calibration
step of regularisation in the image plane was not done as part of our
coherent processing, because the reason for self-calibration was to
increase signal-to-noise on a self-consistent set of visibilities and
closure phases, rather than to form an image.

\begin{figure}
 \centerline{R~Leo at 712\,nm}
 \includegraphics{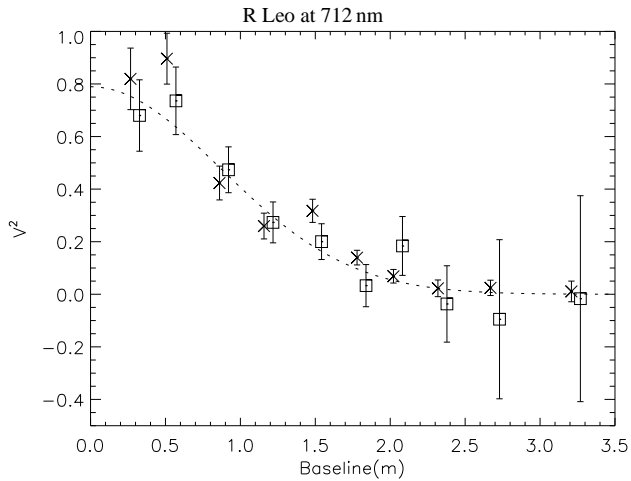}
 \caption{A square-visibility curve for R~Leo in a single
 wavelength channel, obtained by dividing
 the square visibilities of R~Leo by those of its calibrator star,
   $\alpha$-Leo. Crosses represent coherently processed data, squares
   represent conventionally processed (integrated $V^2$) data, and the
   dotted line is the best fit Gaussian.}
 \label{figCoherentEx}
\end{figure}

The final results of coherent averaging are illustrated in Figure
\ref{figCoherentEx}, using R~Leo as an example. It can be seen that the
errors in squared visibility are slightly reduced on short baselines, due
to MAPPIT's wavelength bootstrapping ability, and are greatly reduced for
the longest baselines, due to both wavelength and baseline
bootstrapping. Coherent averaging was used in the processing of all data
taken with the 5-hole mask apart from $\gamma$~Cru, which was in the
photon-rich regime for all wavelengths channels.  We also note that we found
no evidence of any systematic bias when coherently averaged data were
compared to incoherently averaged data.

\section{Results and Discussion}

With up to 40 wavelength channels for each star (note that these are not in
general independent; see Section~\ref{sectDiams}), the raw visibility data
are too extensive to be presented in full. However, an object that we
observed at one position angle to be only partially resolved, or one that
gave poor signal-to-noise at long baselines, could be adequately fitted by
a simple model such as a uniform disk. We thus divided the visibility data
into three categories: under-resolved or poor signal-to-noise data, where
only an estimate of size could be made; data where an estimate of size
could be made and constraints could be placed on the functional form of the
brightness profile; and data with high signal-to-noise, where more free
parameters were required.

\subsection{Rejection of Uniform Disk Model}

\begin{figure}
\centerline{$L_2$ Pup at 820\,nm}
\includegraphics{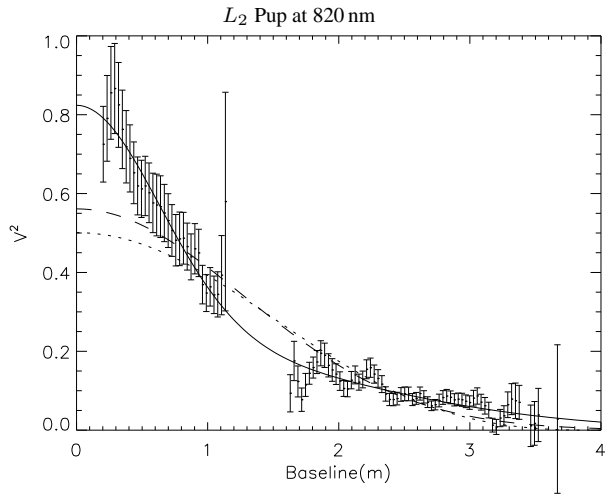}
\caption{The error bars represent calibrated $V^2$ from $L_2$ Pup
  slit data. The gap in $V^2$ at the shortest baselines is due to the
  difficulty in calibration as these baselines are highly seeing
  dependent. The gap between 1.2 and 1.6~m is due to the large central
  obstruction in the AAT preventing sampling of these baselines. The
  dashed line is the best single-component Gaussian
  fit and the dotted line the best uniform disk fit. The solid line
  is a two-component Gaussian fit with FWHMs $25 \pm 2$ mas and $76
  \pm 6$ mas, with the smaller component containing $53 \pm 5\%$ of
  the flux. Other models with a central bright object and extended
  flux could also fit the data.}
\label{figL2Pup2Comp}
\end{figure}

The model of a uniformly illuminated disk has often been the first choice
when fitting to interferometric data.  However, this model can be rejected
for much of the data presented here.  The most striking example of a star
that does not match the uniform-disk model is L$_2$~Pup. For this star, a
two-component Gaussian was the simplest model that fitted the data, as
shown in Figure \ref{figL2Pup2Comp}.

\begin{figure}
\centerline{R~Dor at 847\,nm}
\includegraphics{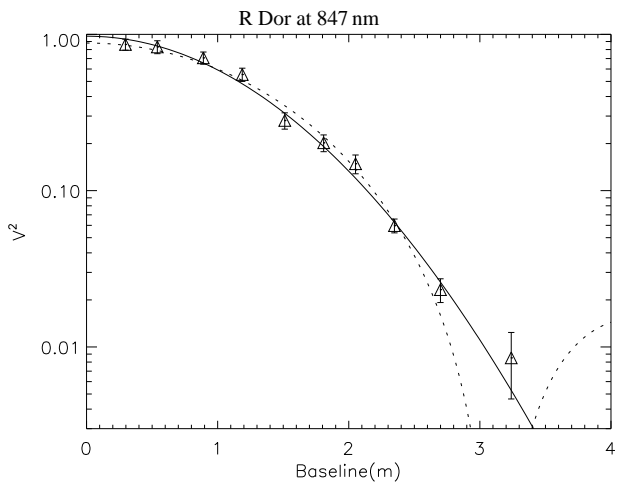}
\caption{Incoherently processed visibilities on a logarithmic scale for 
  R~Dor at 847\,nm, a
  uniform disk model (dotted line) with a reduced $\chi^2$ of 2.19 and
  a Gaussian disk model (solid line) with a reduced $\chi^2$ of
  1.01. Nearby wavelength channels have similar visibility curves.}
\label{figRDorVisplot}
\end{figure}

For W~Hya, R~Dor, $o$~Cet and R~Leo, there were certain wavelength bands
where the uniform disk model could be rejected at a 2~$\sigma$ level, as
given in Table~\ref{tblGaussian}. An example visibility curve is given in
Figure~\ref{figRDorVisplot}. This kind of profile is indicative of an
intensity profile with broad wings. With the exception of the coherently
processed data for R~Dor and R~Leo in strong absorption bands, a Gaussian
brightness profile gave a reasonable fit to all visibilities at all
wavelengths. There were no data sets where the uniform disk profile was
preferred over the Gaussian.  We note, however, that many stars were
under-resolved at certain wavelengths, especially in near-continuum bands,
and many alternative functional forms for the brightness profile (for
example, fully-darkened disks) might also fit as well as the Gaussian. For
a discussion of this, see \citet{Scholz03}.  In the partially-resolved
case, one can multiply by a factor of 1.6 to get an equivalent uniform disk
diameter from a Gaussian FWHM.  In this case, for a bounded intensity
distribution, MAPPIT is only capable of measuring the curvature of the
$V^2$ versus baseline curve near the origin, which corresponds to the
second moment of the intensity distribution. With all this in mind, the
word `diameter' will be used to describe MAPPIT data in the following
sections with the meaning of Gaussian FWHM fitted to available baselines.
To simplify model fitting, data at baselines beyond where $V^2$ first fell
below $0.01$ were ignored.

\begin{table}
\caption{Summary of visibility curve constraints for sample stars.
  With the exception of $L_2$ Pup, a Gaussian disk model provided a
  good fit to $V^2$ in regions where the uniform disk was rejected.}
\begin{tabular}{@{}lrr@{}}
  \hline
   Name     & Uniform disk         & Evidence for \\
            & rejection range (nm) & Asymmetries?\\
 \hline
   $o$~Cet  & 730--850 & No \\  
   R~Hya    &    ---   & No \\
   $\gamma$~Cru &    ---   & No \\
   R~Car    &    ---   & No \\
   R~Leo    & 735--830 & Yes \\
   R~Dor    & 720--900 & Yes \\
   W~Hya    & 720--900 & No \\
   L$_2$~Pup & All (680--920) & Yes \\
\hline
\end{tabular}
\label{tblGaussian}
\end{table}

\subsection{Wavelength Dependent Diameters}
\label{sectDiams}

Given that we have now chosen to fit a single Gaussian profile where
possible (i.e., for all stars except L$_2$~Pup), the visibility data can be 
distilled to give a plot of
the FWHM of the Gaussian profile versus wavelength. 
Such plots are presented in Figures
\ref{figGamCruDiam}--\ref{figL2PupDiam}.
The measured spectrum is overlayed as a dashed line 
(uncorrected for atmospheric absorption, and in arbitrary units)
allowing the position of the absorption bands to be seen clearly
and registered against variations in the apparent size. 
The spectral bandpass of wavelength channels is given as a solid 
line of arbitrary height, (appearing as a trapezoid to the lower right). 
The wavelength scale has been plotted in a non-linear fashion so that the
size and shape of this spectral bandpass remains the same for all 
wavelengths. Note that for the narrow holes/slit (for example, in
Figure~\ref{figGamCruDiam}) each wavelength channel is nearly
independent from its neighbors. Although the MAPPIT instrument was sensitive to
wavelengths between about 660\,nm and 1020\,nm, fits are often not plotted at the shortest
wavelengths when the signal-to-noise ratio was too low to give
a good fit. Data at wavelengths longer than 940\,nm are not plotted
whenever intermittent vignetting caused problems with calibration. 
Results and discussion for individual stars are given below. For a
recent review of previous observations of many of these stars, see
\citet{Scholz03}.

\subsubsection{$\gamma$~Cru}

\begin{figure}
\centerline{$\gamma$~Cru}
\includegraphics{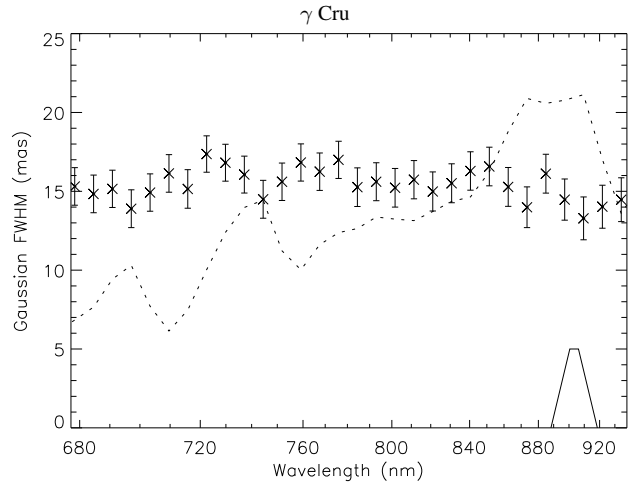}
\caption{FWHM for $\gamma$~Cru of best-fitting Gaussian disk model as a 
function of observing wavelength for $\gamma$~Cru.
The over-plotted dashed line gives the spectrum of the star
(arbitrary units, not corrected for atmospheric absorption),
illustrating for example the well-known TiO absorption 
features at 710 and 760\,nm.
The trapezoidal solid line to the lower right shows the
spectral bandpass.}
\label{figGamCruDiam}
\end{figure}

$\gamma$~Cru is the closest M giant, with spectral type M3.5.  Radial
velocity observations by \citet{Cummings99} indicate a pulsation period of
13--16\,d, and it is probably on the RGB rather than the AGB
\citep{kiss03}.  As can be seen from Figure~\ref{figGamCruDiam}, the
variation of angular diameter with wavelength is not statistically
significant.  Furthermore, the star is barely resolved, so it is meaningful
to quote an equivalent mean uniform disk diameter of $25 \pm 2$ mas. This
agrees with a preliminary diameter at 2.2\,$\mu$m from the VLTI of $24.7
\pm 0.35$ mas \citep{Glindemann01}. The MAPPIT diameter should be viewed
cautiously, however, as the calibrator, $\alpha^1$~Cru, is a single-lined
spectroscopic binary. If the secondary in this system is of near-equal
brightness, then the MAPPIT result could be biased by as much as 4 mas
using the orbit from \citet{Thackeray80}, giving a corrected angular
diameter as high as 29 mas. However, this is unlikely, as both the
brightness ratio and the orbital parameters would have to be `unlucky'.

We may compare our measured diameter with expectations from the 
recently calibrated effective temperature scales of late type stars
which may be found in \citet{vanBelle99a} or \citet{Dumm98}.  Both
give $R_* = 84 R_\odot$ with an error of approximately 8\% due to
the natural dispersion in the relationship. Taking the {\em Hipparcos}
parallax from Table~\ref{tblStarParams} gives an expected diameter of
$28.9 \pm 2.4$ mas, in agreement with our measurement.


%

\subsubsection{$o$~Cet}

\begin{figure}
\centerline{$o$~Cet}
\includegraphics{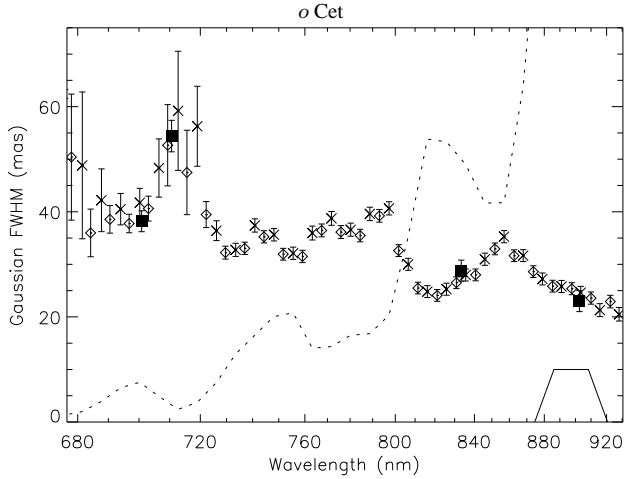}
\caption{Same as Figure~\ref{figGamCruDiam}, but for $o$~Cet.
Crosses and diamonds are separate MAPPIT runs at the same
position angle, at a pulsation phase of 0.38. These data were taken
with a slit mask; additional data taken with
a 5-hole mask (not presented) had lower signal-to-noise but were also consistent 
with the slit data. The four squares are 1993
data from the WHT \protect\citep{Tuthill95}, at a pulsation
phase of 0.48.}
\label{figMiraDiam}
\end{figure}

There are numerous measurements of the diameter of $o$~Cet in the
literature, with the smallest Gaussian FWHM being 12.5 mas in the $J$ Band
\citep{Ireland03}. Previous data from the William Herschel Telescope (WHT)
show excellent agreement with the MAPPIT data, as shown in
Figure~\ref{figMiraDiam}, although the WHT experiment also measured
long-term variations in the diameter of $o$~Cet in their 700 nm and 710\,nm
filters \citep{Tuthill95}. Two clear features in Figure~\ref{figMiraDiam}
are the increase in apparent size in the absorption bands around 710 and
790\,nm by a factor of approximately 1.6 relative to the nearby continuum,
and an overall increase in apparent size toward the blue, with a 1.8 times
increase in apparent diameter at 700\,nm relative to 920\,nm.

\subsubsection{R~Hya}

\begin{figure}
\centerline{R~Hya}
\includegraphics{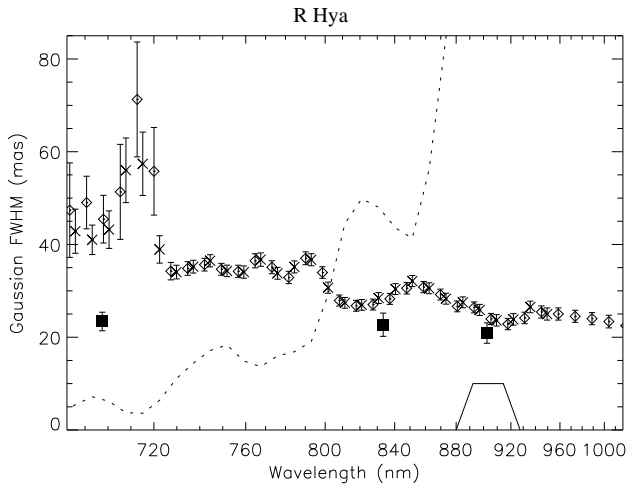}
\caption{Same as Figure~\ref{figGamCruDiam}, but for R~Hya.
Crosses and diamonds are separate MAPPIT runs at position angles
of 72$^\circ$ and 123$^{\circ}$.  Again, the three squares are 1993 WHT data.  The
pulsation phases are 0.28 and 0.62 for the WHT and MAPPIT data,
respectively. }
\label{figRHyaDiam}
\end{figure}

Figure~\ref{figRHyaDiam} again compares MAPPIT wavelength-dependent
diameters to results from the WHT experiment \protect\citep{Haniff95}. The large difference in these data sets could be due to the
difference in pulsation phase and/or long term variability in size between
the two observations. Note that although the apparent size of this
star changes less than $o$~Cet in TiO absorption bands, there is a
larger increase in apparent diameter toward the blue (about a factor of 2.0 between
700\,nm and 920\,nm).

\subsubsection{R~Car}

\begin{figure}
\centerline{R~Car}
\includegraphics{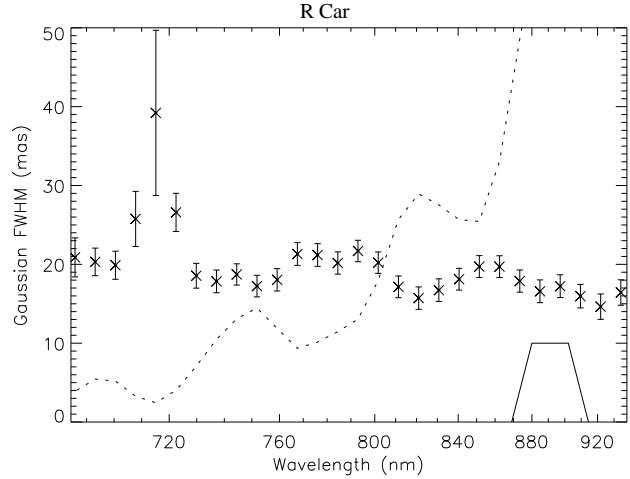}
\caption{Same as Figure~\ref{figGamCruDiam}, but for R~Car.}
\label{figRCarDiam}
\end{figure}

Figure~\ref{figRCarDiam} represents the first published angular diameter
measurement for this star. Although the signal-to-noise is lower than for
the other stars, the increase in apparent size in the 712\,nm TiO band is
clear, as is the relatively small increase in apparent size toward the blue
(roughly a factor of 1.4 between 700\,nm and 920\,nm).

\subsubsection{R~Leo}

\begin{figure}
\centerline{R~Leo}
\includegraphics{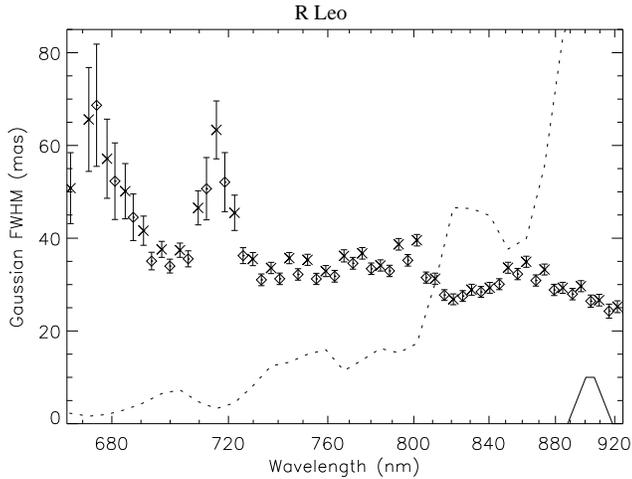}
\caption{Same as Figure~\ref{figGamCruDiam}, but for R~Leo. Crosses and
  diamonds represent data at position angles of $305^{\circ}$ and
  $209^{\circ}$, respectively, using a slit mask. }
\label{figRLeoSlit}
\end{figure}

\begin{figure}
\centerline{R~Leo}
\includegraphics{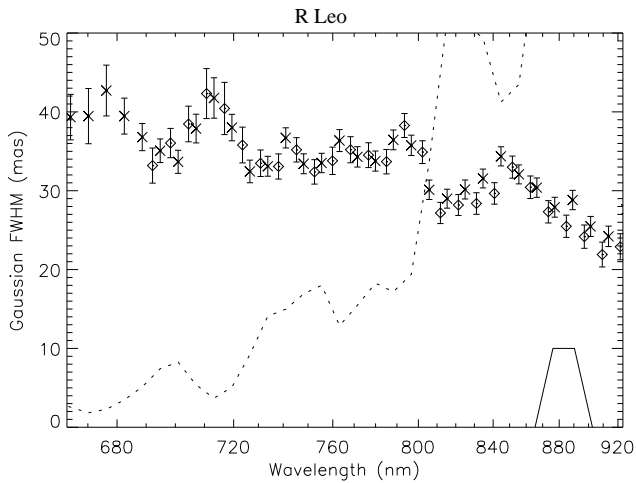}
\caption{Same as Figure~\ref{figRLeoSlit}, but for a hole mask. Crosses and
  diamonds represent data at position angles of $276^{\circ}$ and
  $280^{\circ}$, respectively. The data points in the absorption bands at
  670 and 712\,nm are not valid, because a Gaussian disk is a poor fit to
  the data (in Figure~\ref{figRLeoSlit} signal-to-noise was poor at all
  baselines greater than 1\,m). See section \ref{subAsymm} for details.}
\label{figRLeoHole}
\end{figure}

The two diameter vs. wavelength plots for R~Leo, Figures~\ref{figRLeoSlit}
and \ref{figRLeoHole} show data from a slit mask and a hole mask
respectively. Although these show some differences in detail, as we explain
below, the two curves are consistent within errors. The higher spectral
resolution of the narrow slit enabled a larger increase in apparent
diameter to be seen in the 712\,nm absorption band (roughly a factor of 1.9
larger than nearby continuum). For all other Mira-like stars observed here
with wider slits/holes, the 712\,nm absorption band was slightly
contaminated by neighboring continuum, lessening the change in the apparent
size.  Although the two position angles appear to give different sizes in
the wavelength range 730--850\,nm, note that the data points at neighboring
wavelengths are not independent because they are equally affected by
calibration errors caused by changes in seeing. For the data taken with the
5-hole mask, the smaller apparent change in diameter in the 712\,nm TiO
band is due to two reasons: the effective filter profile was wider than the
slit mask, and the Gaussian profile was a poor fit to the 670 and 712\,nm
data. This poor fit was due to the higher signal-to-noise on the longer
baselines enabled by coherent processing, which gave a two-component-like
fit. The slit data had lower long-baseline signal-to-noise, and were thus
sensitive only to the larger component. The brightness profile
reconstructions in Section~\ref{subAsymm} give a qualitative picture of
this two-component-like appearance.

\citet{Hofmann01} measured the azimuthally averaged Gaussian FWHM for
R~Leo, at a pulsation phase of 0.2, to be $30.5 \pm 1.5$ mas and $34.6 \pm
1.6$ mas at wavelengths of 754 and 781\,nm respectively, consistent within
errors with the MAPPIT result. Their measured FWHM of $23.6 \pm 2.6$ mas at
1045\,nm is consistent with the MAPPIT result at 920\,nm. \citet{Burns98}
measured the diameter of R~Leo at at 830 and 940 \,nm increasing between
phases of 0.1 and 0.6, with measurements at a phase of 0.2 being most
consistent with the MAPPIT diameters (at phase 0.91).

\subsubsection{R~Dor}

\begin{figure}
\centerline{R~Dor}
\includegraphics{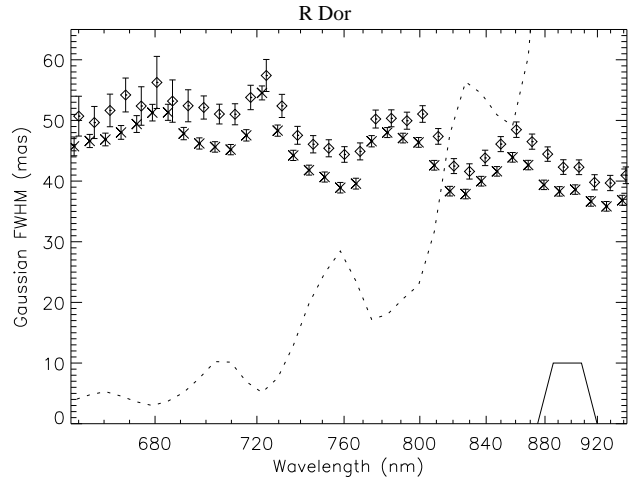}
\caption{Same as Figure~\ref{figGamCruDiam}, but for R~Dor. Crosses and
  diamonds represent data at position angles of $175^{\circ}$ and
  $260^{\circ}$ respectively.}
\label{figRDorDiam}
\end{figure}

R~Dor is a semi-regular variable with a dominant pulsation period
of 332 days, typical of a Mira-like variable, and a secondary
pulsation period of 175 days \citep{Bedding98}. Its diameter has
previously been measured by \citet{Bedding97}, whose results at 1250\,nm
can be converted to a Gaussian FWHM of 36
mas, consistent with the MAPPIT measurements here. The diameter
vs. wavelength 
plot for R~Dor shown in Figure~\ref{figRDorDiam} is noticeably different from those 
for other stars shown in this paper. Both the increase in apparent size toward the blue and the
variation in apparent size in absorption bands are
considerably less than for the large-amplitude Mira stars
presented thus far. Furthermore, the data from the two position angles
separated by approximately 90 degrees show significantly different
apparent sizes. Fortunately, this star is well-resolved and the details
of this apparent asymmetry are investigated further in Section~\ref{subAsymm}.

\subsubsection{W~Hya}

\begin{figure}
\centerline{W~Hya}
\includegraphics{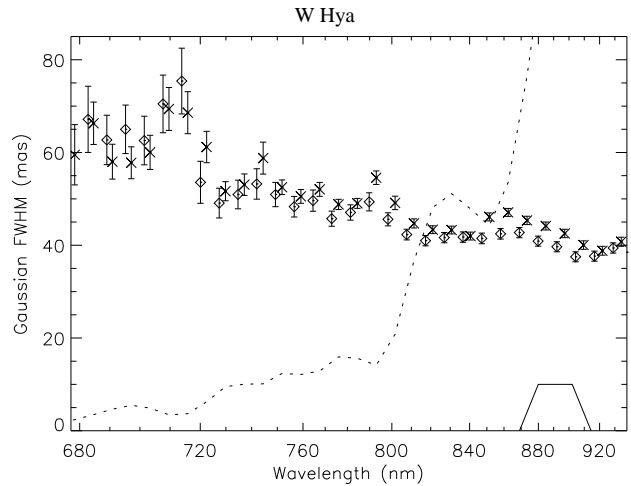}
\caption{Same as Figure~\ref{figGamCruDiam}, but for W~Hya. Crosses and
  diamonds represent data at position angles of $252^{\circ}$ and
  $120^{\circ}$ respectively. }
\label{figWHyaDiam}
\end{figure}

W~Hya is also classified as a semi-regular variable, but has only
one dominant pulsation period of 361 days and has a visual amplitude
of 3 magnitudes, meaning it should be classified as a Mira
variable. Its diameter vs. wavelength plot in Figure~\ref{figWHyaDiam}
has less variation of apparent diameter in absorption bands than the other
large-amplitude Mira stars presented here, but shows the same strong
increase in apparent diameter toward the blue (roughly a factor of
1.6 between 920 and 700\,nm). Although there are some signs of
asymmetries at the 10\% level around 850\,nm, these cannot be
interpreted further due to the  unavailability of phase information,
as phase information recovered from slit masks has a low signal-to-noise ratio.

\subsubsection{L$_2$~Pup}

\begin{figure}
\centerline{L$_2$~Pup}
\includegraphics{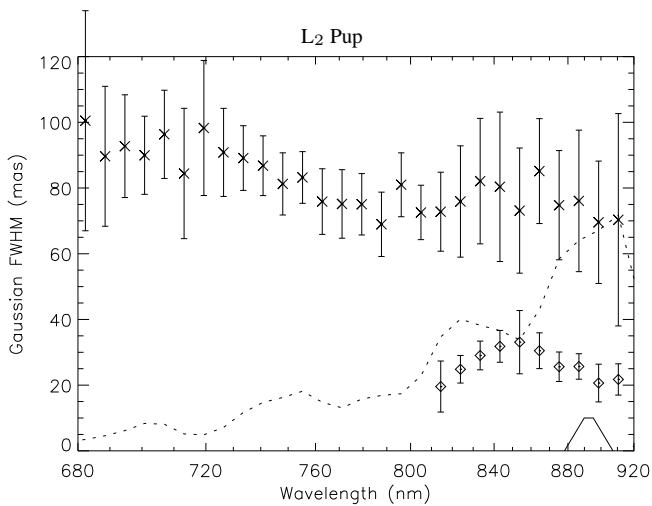}
\caption{Same as Figure~\ref{figGamCruDiam}, but for L$_2$~Pup, for data
  taken with a slit mask. Crosses represent the larger of two components in
  a two-component fit, and diamonds represent the smaller component. Long
  baseline signal-to-noise was too low at wavelengths shorter than 800\,nm
  for the smaller of the two components to be significantly resolved.}
\label{figL2PupDiam}
\end{figure}

L$_2$~Pup is a semi-regular variable with a pulsation period of 140 days
and no previous measurements of its diameter. Using the angular size versus
$V - K$ colour relationship for variable stars from \citet{vanBelle99a},
taking the mean $K$ magnitude of $-$2.24 from \citet{Whitelock00} and a $V$
magnitude of 5.2, the predicted Gaussian FWHM of L$_2$~Pup is 15 mas. This
is consistent with the smaller of the two components shown in
Figure~\ref{figL2PupDiam}. The origin of the larger component will be
discussed further in Section \ref{subAsymm}.


\subsubsection{Implications for Mira atmospheres}

The five Mira stars share three common features: Firstly,
   there is an enlargement by up to a factor of 2 in the strongest TiO bands at
   712 and 670\,nm, when compared to diameters at nearby
   wavelengths. Contamination by nearby continuum radiation is a
   significant factor for all these data, except for R~Leo data
   taken with the narrow slit mask. In particular, the ``true''
   apparent enlargement of R~Hya and $o$~Cet in the 710\,nm
   absorption band would roughly be a factor of 2.0 rather then 1.4
   and 1.6 as shown above, if a narrower filter were used. This kind of
   extension is expected from dynamical models such as those
   investigated in \citet{Jacob02}. Note that the maximum extension
   produced by hydrostatic models is around 30\% over the wavelength
   range considered here \citep{Hofmann98b}, demonstrating the need
   for dynamical models in modeling Mira-like atmospheres.

\begin{figure}
\centerline{All Stars}
\includegraphics{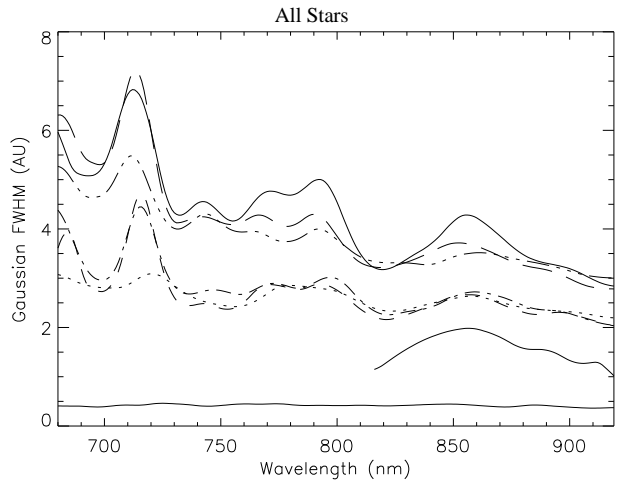}
\caption{Linear size vs.\ wavelength for all stars in the sample, using
their {\em Hipparcos} parallaxes (given in
Table~\ref{tblStarParams}). Diameters for different runs on the same star
have been averaged together and smoothed. From top to bottom the stars are
$o$~Cet (solid line), R~Hya (large dashes), W~Hya (triple-dot dashed), R~Leo
(dot-dashed), R~Car (small dashes), R~Dor (dotted), the smaller
component of L$_2$~Pup (solid line) and $\gamma$~Cru (solid line).  The
larger component of L$_2$~Pup (not shown) has an average size of 5~AU.}
\label{figAllDiam}
\end{figure}

Secondly, the minimum diameters of these stars are systematically too large
   for published models. 
   The most extreme dynamical fundamental-mode models from
   \citet{Jacob02} predict a maximum 920\,nm radius of $340
   R_\odot$ for a short to medium baseline fit to their models, which
   corresponds to a FWHM of 2.0~AU in Figure~\ref{figAllDiam}. This
   means that three stars in this group are systematically
   too large for all of those Mira-like models. 
   Hydrostatic models such
   as those by \citet{Bessell91} are considerably smaller. Although models
   that pulsate in the first overtone can explain these larger sizes,
   they cannot explain the small infra-red sizes observed (for
   example, see \citealt{Ireland03} or \citealt{Perrin99}). It has been
   pointed out in the model study of \citet{Bedding01} that
   scattering by dust in the upper
   atmosphere of Miras could explain these larger apparent diameters,
   as well as their two-component or Gaussian-like appearance at
   wavelengths shorter than 1 micron. Dust was
   also cited by \citet{Hofmann01} to explain their observations of large apparent
   diameters. \citet{Danchi94}
   measured the size of the dust shells around several Mira-like
   stars, and found an inner radius of the dust shell less than 2.7
   continuum radii from the star's centre, and predicted an optical
   depth (including scattering and absorption) of 0.5 for a 700\,nm
   wavelength. This is enough to explain the larger observed diameters
   as compared with those expected from dust-free model studies such as \citet{Jacob02}.

Finally, there is a trend of increasing apparent size toward the blue, which is
   not always correlated with an increase in apparent size in TiO
   bands. W~Hya is the most striking example of this, with a large
   increase in apparent diameter toward the blue but only small
   changes in apparent diameter in absorption bands. W~Hya and R~Hya
   (which also shows a large increase in apparent size toward the
   blue) are also the stars observed closest to minimum. If
   scattering by dust is a major contributor to this effect, then this
   might suggest that significant dust production occurs near minimum.

\subsection{Asymmetrical Targets}
\label{subAsymm}

\begin{figure}
\centerline{R~Dor}
\includegraphics{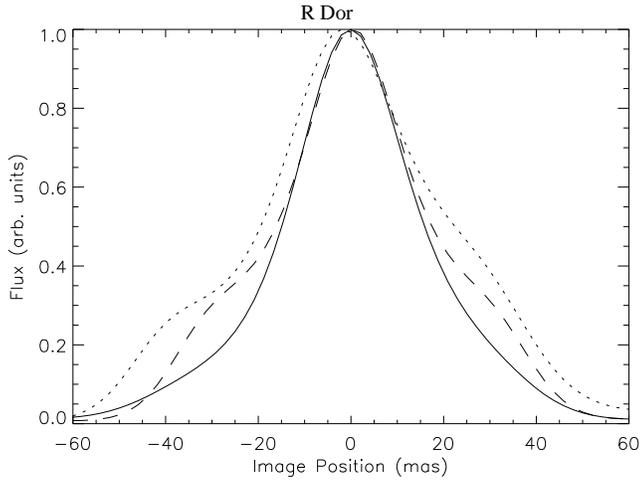}
\caption{MEM reconstructed brightness profiles of R~Dor at a position angle
  of 175$^\circ$. The solid line is at 750\,nm, the dotted line at 712\,nm
  and the dashed line at 700\,nm.}
\label{figRDorProfile1}
\end{figure}

\begin{figure}
\centerline{R~Dor}
\includegraphics{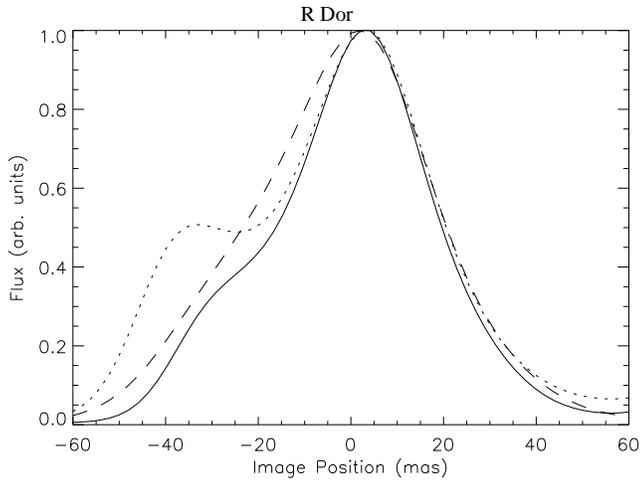}
\caption{The same as Figure~\ref{figRDorProfile1}, but at a position angle of
  260$^\circ$.}
\label{figRDorProfile2}
\end{figure}

\begin{figure}
\centerline{R~Dor}
\includegraphics{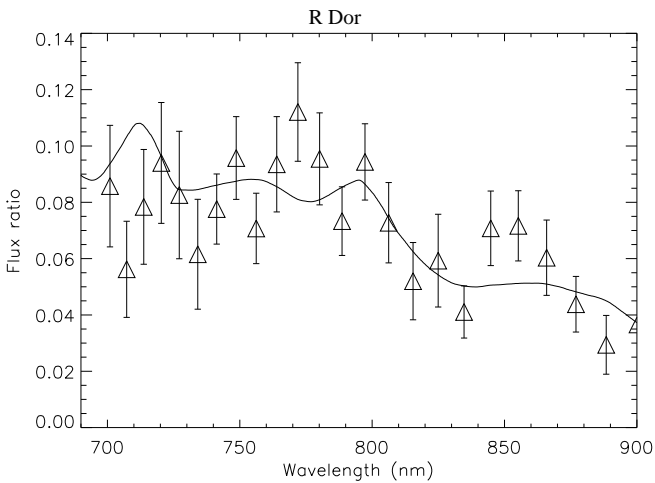}
\caption{The ratio of point source to disk flux for the model of R~Dor
  described in the text, overlaid with a scaled ratio of an M6 to an M8
  model spectrum \protect\citep{Fluks94}.}
\label{figRDorFit}
\end{figure}

\begin{figure}
\centerline{R~Leo}
\includegraphics{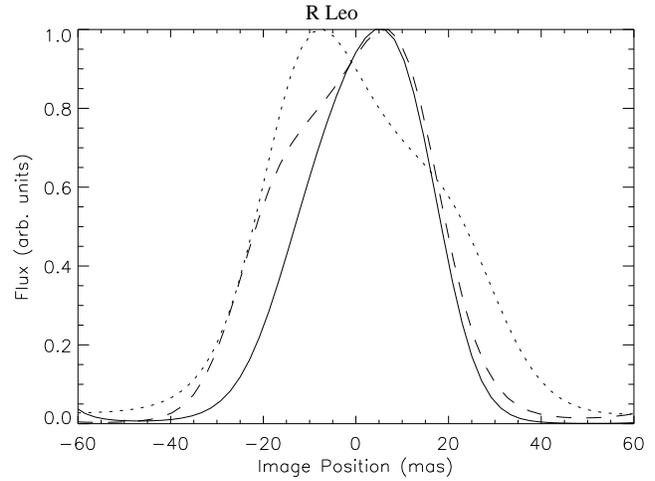}
\caption{MEM reconstructed brightness profiles of R~Leo at a position angle
  of 276$^\circ$. The solid line is at 815\,nm, the dashed line at 700\,nm
  and the dotted line at 712\,nm.}
\label{figRLeoProfile}
\end{figure}

\begin{figure}
\centerline{L$_2$~Pup}
\includegraphics{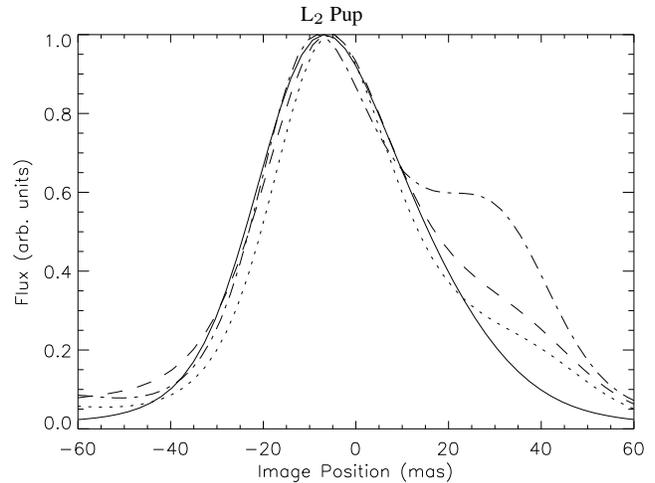}
\caption{MEM reconstructed brightness profiles of L$_2$~Pup at a position
angle of 178$^\circ$. The
  solid line is at 850\,nm (weak TiO), the dotted line at 820\,nm
  (pseudo-continuum), the dashed line at 790\,nm (weak TiO) and the
  dot-dashed line at 750\,nm (pseudo-continuum).}
\label{figL2PupProfile}
\end{figure}

We see asymmetries, as indicated by non-zero closure phases,
in the stars R~Leo, R~Dor and L$_2$~Pup. The maximum closure
phase observed was $-90^\circ$ at 5~$\sigma$ for R~Dor. For these stars, which
also have reasonable signal-to-noise at visibilities less than
0.1, a simple model diameter
does not fully represent the available data. We thus 
present reconstructed one-dimensional brightness profiles. 
Data at a variety of position angles would be needed in order to
reconstruct a two-dimensional image, but the one-dimensional data
still allows quantification of the wavelength-dependent structure in
these stars. The one-dimensional profiles were constructed using the
Maximum Entropy (MEM) algorithm as described in
\citet{Wilczek85}. Each point in the profile can be thought of as
the integral of the image flux along an axis perpendicular to the
position angle of observation. Flux is scaled so that the peak flux is
always 1.0 in order to easily show the difference in the shapes of the
flux distributions. The absolute positions of the profiles are also
arbitrary, as uncertainties in the dispersion of the optical system made
differential phase measurements impossible over the full
bandwidth. 

The three stars presented here have quite different characteristics. The
brightness profiles of R~Dor in Figures~\ref{figRDorProfile1} and
\ref{figRDorProfile2} are dominated by a central Gaussian-like profile with
additional features that become stronger in the absorption bands. For the
observations at a position angle of $260^\circ$, it was possible to model the flux as a
central Gaussian disk with FWHM that varied with wavelength, and a point
source offset by $36.2 \pm 0.5$~mas in all wavelength channels. The ratio
of the flux in these two components at wavelengths between 700 and 900\,nm
is given in Figure~\ref{figRDorFit}, along with a tentative fit based on a
temperature difference between components. It is clear that the point
source in the model contributes to more of the total flux toward the blue,
however it is not clear whether this is due to a spatial variation in
temperature causing a spatial variation in TiO opacity, or from an effect
in the inner circumstellar environment such as scattering by dust.

The brightness profiles of R~Leo in Figure~\ref{figRLeoProfile} have
opposite directions of skew in the 712\,nm absorption band and the
continuum, suggesting that the flux in this band is dominated by one or
more non-central bright spots. Quantitative interpretation of this is
difficult, because the phase signal is weak, so model-fitting the flux in
two components: a central disk and off-centre spot, gives a large error in
the relative flux in the two features at all wavelengths. The strength of
these asymmetries is similar to that found in earlier work by
\citet{Tuthill99}.

The asymmetry in the profiles of L$_2$~Pup increases toward the blue, as
shown in Figure~\ref{figL2PupProfile}. There is no significant increase in
the strength of the asymmetrical feature in the weak absorption bands at
790 and 850\,nm. The signal-to-noise for shorter wavelengths than those
presented here was low, but the trend in the two component fits from
section \ref{sectDiams} was that the central component was contributing
relatively less flux toward the blue. We therefore interpret the wide
asymmetrical feature as scattering by dust in the circumstellar
environment. The large change in the strength of this feature between 750
and 850\,nm relative to the central source is indicative of the extent of
dust extinction for this star. This extinction had increased by
approximately 2 magnitudes in $V$ immediately prior to these observations
\citep{Bedding02}.

\section{Conclusion}

Simultaneous spatial and spectral data have been recorded
for a sample of eight red giants using the novel
cross-dispersed interferometry technique developed by 
the MAPPIT project.
With spectral resolving power $\lambda/\delta\lambda< 100$ and spatial resolutions
sufficient to measure structures $> 10$\,mas in size, it has 
been possible to recover the angular size distribution as a 
function of wavelength spanning the R and I-bands for our sample stars. A
combination of wavelength and baseline bootstrapping enabled coherent
averaging of many data sets, enabling a large increase in the
signal-to-noise ratio.
With the exception of $\gamma$~Cru, which had a much earlier
spectral type (M3.5) than the rest of the sample, diameters
were found to exhibit dramatic changes with wavelength.
These diameter excursions occurred both as broad trends
with wavelength across the entire band, and in narrow 
spectral windows. 
In the latter case, the changes were manifest as enlargements
across spectral regions associated with strong TiO absorption,
although these were mixed with nearby quasi-continuum layers
at the spectral resolutions achieved here, arguing for further
work with higher dispersion.

For the Mira variables in the sample, Gaussian-like (or 
other tapered) radial profiles were found to give a
better fit to the data than uniform disk profiles. The prevalence of
these Gaussian-like profiles, the increase in apparent size toward the
blue as separate from molecular effects and the larger than expected
apparent sizes even at 920\,nm all point toward the significance of
scattering by dust in the inner circumstellar environment affecting
interferometric observations at these wavelengths.
Non-centro-symmetric elements were detected for 3 stars, 
which may be explained as thermal or opacity inhomogeneities
in the stellar atmosphere or inner circumstellar regions.

The star L$_{2}$~Pup was found to be something of a special 
case, with the visibility data betraying the presence of
two resolved components, which we interpret as a stellar
disk and a dusty circumstellar envelope. 
Asymmetries detected in this star raise the interesting 
possibility of resolved highly clumpy structure as close
as the dust condensation radius.
 
These results are all in accord with the theoretical and
observational picture of pulsating late-M giants which emphasizes the
potential for the extended molecular atmosphere to dramatically
affect the observed properties of the stars.



\section*{Acknowledgments}

Visual data for estimating the phases of the MAPPIT
observations were supplied by the AFOEV, the VSOLJ and Albert Jones. 
We thank M.~Scholz for many valuable discussions, and the
Anglo-Australian Observatory for their support for the duration of the
MAPPIT experiment. This research was in part supported by the
Australian Research Council and
the Deutsche Forschungsgemeinschaft within the linkage project ``Red Giants.''

\bibliography{mnemonic-simple,thesis}
\bibliographystyle{mynatbib}

\label{lastpage}

\end{document}